\documentclass[12pt]{article}
\begin{document}
\bigskip
\hskip 5in\vbox{\baselineskip12pt \hbox{NSF-KITP-04-101
} }
\bigskip
\begin{center}
\Large Finite Density States in\\[1mm] Integrable Conformal Field Theories
\end{center}
\bigskip
\bigskip

\begin{center}
Nelia Mann\\
Department of Physics, \\
University of California, \\
Santa Barbara, CA\ 93106, USA\\
E-mail: nelia@physics.ucsb.edu
\end{center}
\bigskip

\begin{center}
Joseph Polchinski\\
Kavli Institute for Theoretical Physics,\\
University of California, \\
Santa Barbara, CA 93106-4030, USA\\
E-mail: joep@kitp.ucsb.edu
\end{center}
\bigskip
\bigskip

\begin{abstract}{We study states of large charge density in integrable conformal
coset models.
For the $O(2)$ coset, we consider two different S-matrices, one corresponding
to a Thirring mass perturbation and the other to the continuation to
$O(2\!+\!\epsilon)$.  The former leads to simplification in the conformal limit;
the latter gives a more complicated description of the $O(2)$ system, with a
large zero mode sector in addition to the right- and left-movers. We argue that
for the conformal  $O(2\!+\!2M|2M)$ supergroup coset, the S-matrix is given by the
analog of the $O(2\!+\!\epsilon)$ construction.}
\end{abstract}

\vspace{0.5cm}
\newpage
\tableofcontents
\baselineskip=17.6pt

\section{Introduction}

The $AdS_5 \times S^5$ background of IIB string theory is highly
symmetric, and one might hope that the string world-sheet theory in
this background would be exactly solvable.  However, because of the
presence of R--R flux, the usual tools such as current algebra are not
available.  In Ref.~\cite{MSW,BPR,vallilo,polyakov} it was shown
that the world-sheet CFT has an infinite number of nonlocal conserved
charges of the type that arise in integrable models, at least at the
classical level (see Refs.~\cite{alday,HPXY,HY} for further developments).
There has been some discussion of the combination of integrability and
conformal invariance \cite{ZZ1,FS,BLZ}, but thus far there are no
methods with the power and generality of current algebra or rational conformal
field theory.  Thus, for the $AdS_5 \times S^5$ world-sheet
theory, and more generally for CFT's based on supermanifold sigma
models~\cite{sethi,BVW,BSZ,zirn,RS,SWK}, there is no known way to calculate
the energies of general world-sheet states.

In this paper we would like to take small steps in this direction.
There are well-established methods for calculating the energies of
states with large densities of a conserved charge, starting from the
exact continuum S-matrix~\cite{PW,JNW,has1,has2,FNW,EH}.  We would like to
examine the conformal limit of these calculations, and then apply them to
the conformal $OSp(2\!+\!2M|2M)$ coset model.

In Sec.~2 we develop the conformal limit of the finite density system.
The calculation separates into decoupled right- and left-moving
calculations, which are simpler than in the nonconformal case.  As a
warmup we apply this first to the case $M=0$, the bosonic $O(2)$ model.
In Sec.~3 we use the normal Thirring description of the $O(2)$ model, which has
a simple massless limit.
In Sec.~4
we consider a different description of the $O(2)$ model, as the $N \to
2$ limit of the $O(N)$ model.  This gives a different S-matrix,
describing the bosonic spins of the $O(2)$ model rather than the
Thirring fermions.  The limiting process seems to be sensible but the
result is more complicated than previous examples of conformal
integrable models, in that there is a large and nontrivial zero-mode
sector in addition to the right- and left-movers.  In Sec.~5 we argue
that the $OSp(2\!+\!2M|2M)$ model should be given by the lift of the {\it
second} description of the $O(2)$ model.  Section~6 discusses further
directions.

Of course, there has been an explosion of work on integrability from
the gauge theory side of the AdS/CFT duality, beginning with
Refs.~\cite{MZ,BKS}.  At present, it appears that progress in this
direction is much easier than on the string sigma model side.  However,
it seems likely that a perspective from both sides of the duality will
ultimately be useful.  We should note that states with large spin have
considered extensively on the gauge theory side as well (e.g.
\cite{FT,BMSZ}; for a review see ref.~\cite{arkrev}).  Also, there have
been efforts to derive the string sigma model directly from the spin
chain on the gauge theory side~\cite{sig1,sig2,sig3,sig4,sig5,sig6,swan,AFS} and
to relate the integrable structures on the two sides~\cite{DNW,arustau,eng}; we do
not know if there is a connection with our work.

\section{Finite Density in the Conformal Limit}

We start with a 1+1 dimensional relativistic theory, whose exact
S-matrix is assumed to be known.  We are interested in the the lowest
energy state with a specified charge and momentum, so we consider the
case that we only have particles of one type, with a given sign of the
charge, and that all states are filled in a range of rapidities $-B_L <
\theta < B_R$.  We then have the standard Bethe ansatz
equation~\cite{bethe,yang}:
\begin{equation}
m \cosh \theta + \int_{-B_L}^{B_R} K(\theta - \theta') \rho(\theta')
d\theta'
= \rho(\theta)\ ,\quad -B_L < \theta < B_R\ . \label{eq:bethe}
\end{equation}
Here $\rho$ is $2\pi$ times the density of particles per unit length
{\it and} unit rapidity, so that the number density per unit length is given by the rapidity integral
\begin{equation}
{\mathcal J} = \frac{1}{2\pi} \int_{-B_L}^{B_R} \rho(\theta)
d\theta\ .
\end{equation}
The kernel is
\begin{equation}
K(\theta) = \frac{1}{2\pi i} \,\partial_\theta \ln S(\theta)\ ,
\end{equation}
where $S(\theta)$ is the S-matrix between two particles of the given
type.  Note that the integral equation holds only in the range $-B_L < 
\theta < B_R$ in which $\rho(\theta)$ is nonzero.  The equation is
complicated because this range is bounded on both ends.  It can be
analyzed using the Wiener-Hopf technique~\cite{JNW,has1,has2,EH}, but
in general cannot cannot be solved in closed form.

Now let us take the limit $m \to 0$, holding fixed the momentum.  For
right- and left-moving particles,
\begin{eqnarray}
p_R &=& m \sinh \theta \approx \frac{m}{2}\, e^{\theta} = \frac{\mu}{2}\,
e^{\tilde\theta_R}\ , \quad \theta = \tilde\theta_R + \ln
\frac{\mu}{m}\ ,\nonumber\\[2mm]
p_L &=& m \sinh \theta \approx -\frac{m}{2}\, e^{-\theta} = -\frac{\mu}{2}\,
e^{-\tilde\theta_L}\ , \quad \theta = \tilde\theta_L - \ln
\frac{\mu}{m}\ ,
\end{eqnarray}
where $\mu$ is a fixed reference scale.  Thus we hold fixed
$\tilde\theta_{R,L}$ in the limit.
We assume that in the massless limit the density separates into a
right-moving\pagebreak[3] part which is a function of $\tilde\theta_R$ and a
left-moving part which is a function of $\tilde\theta_L$:
\vspace{-3mm}
\begin{eqnarray}
\rho_R(\tilde\theta_R) &=& \lim_{m\to 0} \rho(\tilde\theta_R + \ln
{\mu/m})\ ,
\nonumber\\[1mm]
\rho_L(\tilde\theta_L) &=& \lim_{m\to 0} \rho(\tilde\theta_L - \ln
{\mu/m})\ . \label{eq:conlim}
\end{eqnarray}

Since the S-matrix depends only on rapidity differences, the RR and LL
S-matrices are
the same as the original S-matrix:
\begin{equation}
S_{RR}(\tilde\theta - \tilde\theta') =
S_{LL}(\tilde\theta - \tilde\theta') =
\lim_{m \to 0} S(\tilde\theta - \tilde\theta') \ .
\end{equation}
On the other hand, for right- and left-moving particles the rapidity
difference is diverging in the limit and so\footnote{
We are assuming here that the limits $m\to 0$ and $\theta \to \infty$ commute.  This will be true for the Thirring S-matrix studied in Sec.~3, which is simply independent of $m$, but it will not be true for the limit in Sec.~4, which will require a more complicated treatment.}
\begin{equation}
S_{RL}(\tilde\theta - \tilde\theta') = \lim_{m \to 0}
\lim_{\theta \to \infty} S(\theta)\ .
\end{equation}
The Bethe ansatz equation then separates into two pieces, which are
obtained by holding $\tilde\theta_R$ or $\tilde\theta_L$ fixed as $m
\to 0$:
\begin{eqnarray}
\frac{\mu}{2}\, e^{\tilde\theta} 
   + \int_{-\infty}^{\tilde B_R} K(\tilde\theta - \tilde\theta')
\rho_R(\tilde\theta') d\tilde\theta'
&=&  \rho_R(\tilde\theta)\ ,\quad -\infty < \tilde\theta < \tilde B_R\ ,
\nonumber\\[1mm]
\frac{\mu}{2}\, e^{-\tilde\theta}
   + \int_{-\tilde B_L}^{\infty} K(\tilde\theta - \tilde\theta')
\rho_L(\tilde\theta') d\tilde\theta'
&=&   \rho_L(\tilde\theta)\ ,\quad -\tilde B_L < \tilde\theta < \infty\ ,
\label{eq:betheconf}
\end{eqnarray}
where $\tilde B_{R,L} = B_{R,L} - \ln \mu/m$ is fixed in the limit.
There is no RL cross term because $\partial_\theta S$ vanishes at large rapidity for all cases of interest.

Because the original rapidity range $B_R + B_L$ diverges in the limit,
the right- and left-moving rapidity ranges are each bounded on only one
side, and these integral equations can be solved in closed form.  We
follow the Wiener-Hopf technique, as described for example in the appendix to
\cite{JNW} and in
\cite{FNW}.  Focussing on the
right-moving equation, we write it as
\begin{equation}
g(\tilde\theta) - \rho_R(\tilde\theta)
   + \int_{-\infty}^{\tilde B_R} K(\tilde\theta - \tilde\theta')
\rho_R(\tilde\theta') d\tilde\theta'
= X(\tilde\theta)\ ,
\end{equation}
where $X(\tilde\theta)$ is nonvanishing only for $\tilde\theta > \tilde
B_R$. Here $g(\tilde\theta) = \frac{1}{2} \mu e^{\tilde\theta} H(
\tilde B_R - \tilde\theta)$, where $H$ denotes the step function.
Taking the Fourier transform $\int_{-\infty}^\infty d\tilde\theta\,
e^{i \omega\tilde\theta}$ on both sides gives
\begin{equation}
\tilde g(\omega) - [1 - \tilde K(\omega) ] \tilde \rho(\omega) = \tilde
X(\omega)\ .  \label{eq:iefour}
\end{equation}
Because of the bounded ranges of $\rho_R$, $g$, and $X$, it follows that
\begin{eqnarray}
\tilde\rho_R(\omega) &=& e^{i \omega \tilde B_R}
\rho_{R-}(\omega)\ , \nonumber\\[1mm]
\tilde g(\omega) &=& e^{i \omega \tilde B_R}  g_-(\omega)\ ,
\nonumber\\[1mm]
\tilde X(\omega) &=& e^{i \omega \tilde B_R} X_{+}(\omega) \ , 
\end{eqnarray}
where the subscripts $\pm$ denote functions which are holomorphic in
the upper and lower half-planes respectively.  These functions also
vanish asymptotically in the half-planes where they are holomorphic,
because $\rho_R$, $g$ and $X$ have finite discontinuities at $\tilde
B_R$.

Given a bounded function $\Psi(\omega)$ which vanishes at $\omega \to
\pm \infty$, we can define
\begin{equation}
\Psi_{\pm}(\omega) =  \pm \frac{1}{2\pi i} \lim_{\delta \to 0}
\int_{-\infty}^\infty \frac{d\omega'\,\Psi(\omega')}{\omega' -
(\omega\pm i \delta)}\ . \label{eq:pmproj}
\end{equation}
These are holomorphic in the upper and lower half-plane respectively,
and moreover
$\Psi(\omega) = \Psi_+(\omega) + \Psi_-(\omega)$.  The operations $[\ \
]_{\pm}$ act as projection operators, in that $[f_-]_+ = 0$ and
$[f_-]_- = f_-$.
Applying this to $\ln[1-\tilde K(\omega)]$ (which vanishes
asymptotically for smooth $K(\theta)$),
it follows that we can write
\begin{equation}
1 - \tilde K(\omega)  = \frac{1}{G_+(\omega) G_-(\omega)}\ ,
\end{equation}
where $G_\pm(\omega)$ are holomorphic and nonvanishing in the upper and
lower half-planes respectively, and approach 1 asymptotically.
The integral equation can thus be put in the form
\begin{equation}
\frac{ \rho_{R-}(\omega) }{G_-(\omega)} =
G_{+}  g_-(\omega) - G_+(\omega)  X_+(\omega)\ .
\end{equation}
Taking the $[\ \ ]_-$ part~(\ref{eq:pmproj}) eliminates the unknown
function $X_+(\omega)$,\,\footnote{\,This is the point where the
simplification due to a semi-infinite range enters.  Otherwise there
would be a second unknown function $ X_-(\omega)$, and an
additional step would be needed, leading to an integral equation that cannot be solved in closed form.} giving
\begin{equation}
\frac{ \rho_{R-}(\omega) }{G_-(\omega)} =
[ G_{+}  g_-(\omega) ]_-\ .
\end{equation}
Finally, using the explicit form $ g_-(\omega) = \mu e^{\tilde
B_R}/2(1+i\omega)$ allows us to evaluate the contour integral
explicitly, giving
\begin{equation}
{\tilde \rho_{R}(\omega) } = \frac{\mu\, e^{(1+i\omega)\tilde B_R} G_+(i)
G_-(\omega)}{2( 1 + i\omega)}\ .
\end{equation}

The rapidity density $\rho_R(\theta)$ is obtained from the inverse Fourier
transform, but the quantities of main interest are given directly by
$\tilde \rho_{R}(\omega)$.
The total charge density carried by the right-movers is
\begin{eqnarray}
{\mathcal J}_R = \frac{1}{2\pi} \int_{-\infty}^\infty
\rho_{R}(\tilde\theta)\, d\tilde\theta &=& \frac{1}{2\pi} \,\tilde
\rho_{R}(0) \nonumber\\
&=& \frac{\mu  e^{\tilde B_R}}{4\pi}\, G_+(i) G_-(0)\ .
\end{eqnarray}
This determines $\tilde B_R$ in terms of ${\mathcal J}_R$.
The total energy and momentum densities are
\begin{eqnarray}
{\mathcal E} = {\mathcal P} =  \frac{\mu}{4\pi}
\int_{-\infty}^\infty e^{\tilde\theta} \rho_{R}(\tilde\theta)\,
d\tilde\theta &=& \frac{\mu}{4\pi} \,\tilde \rho_{R}(-i)
\nonumber\\
&=& \frac{\mu^2  e^{2\tilde B_R}}{16 \pi}\, G_+(i) G_-(-i)\ .
\end{eqnarray}
In general $K(\theta) = K(-\theta)$, and so $G_+(i) =
G_-(-i)$.  Then we can write
\begin{equation}
{\mathcal E} = {\mathcal P} =  \frac{\pi {\mathcal
J}_R^2}{G_+(0)G_-(0)}
= [1 - \tilde K(0)] \pi {\mathcal J}_R^2\ . \label{eq:ep1}
\end{equation}
Similarly, for left-movers
\begin{equation}
{\mathcal E}= -{\mathcal P}
= [1 - \tilde K(0)] \pi {\mathcal J}_L^2\ . \label{eq:ep2}
\end{equation}
The relation between the energy and charge thus depends only on the total
change
in the phase of $S$ from $\theta = -\infty$ to $\theta=\infty$.

Note that we have discussed only densities in a system of infinite
volume.  In a finite volume system there will be corrections, Casimir
effects.  Obtaining these from the infinite volume S-matrix which is
our starting point is a difficult problem for which there is only a
partial solution; we will comment on this further in the conclusions.
For the bulk of this paper we focus on the infinite volume case, or
equivalently on the {\it leading} high-density properties in a system
of finite volume.

\section{The Massless Thirring Model}

The fermionic and bosonic descriptions of the Thirring model
are
\begin{equation}
S = \int d^2 x\, \biggl[ i \bar\psi \gamma^\mu \partial_\mu\psi +
\frac{\lambda}{2} (\bar\psi\psi)^2 - m \bar\psi\psi
\biggr]
\end{equation}
and
\begin{equation}
S = - \int d^2 x\, \biggl[ \frac{1}{2g^2}\partial_\mu \phi \partial^\mu
\phi + m \cos\phi \biggr]\ . \label{bosact}
\end{equation}
The field $\phi$ is normalized to have periodicity $2\pi$, so that a 
fermion corresponds to a
kink $\Delta\phi = 2\pi$.  The Thirring fermion-fermion S-matrix
is~\cite{ZZS}
\begin{equation}
S = \frac{\Gamma\Bigl( \frac{8\pi}{\gamma} \Bigr)
\Gamma\Bigl(1 +\frac{8i\theta}{\gamma} \Bigr)
}{\Gamma\Bigl(\frac{8\pi}{\gamma}+\frac{8i\theta}{\gamma}  \Bigr)}
\prod_{n=1}^\infty
\frac{R_n(\theta) R_n(i\pi - \theta) }{R_n(0) R_n(i\pi) } 
\label{eq:thirs}
\end{equation}
where
\begin{equation}
R_n(\theta) =
\frac{\Gamma\Bigl(2n\frac{8\pi}{\gamma}+\frac{8i\theta}{\gamma}  \Bigr)
\Gamma\Bigl(1 + 2n\frac{8\pi}{\gamma}+\frac{8i\theta}{\gamma}
\Bigr)}{\Gamma\Bigl( [2n+1]\frac{8\pi}{\gamma}+\frac{8i\theta}{\gamma}
\Bigr)\Gamma\Bigl(1 + [2n-1]\frac{8\pi}{\gamma}+\frac{8i\theta}{\gamma}
   \Bigr)}\ .
\end{equation}
Only when the Thirring mass is nonzero can this be interpreted as an
S-matrix in the usual sense, but even in the massless limit it can be
used sensibly in the Bethe ansatz~\cite{ZZ1}.  The S-matrix contains a
dimensionless parameter $\gamma$ which is related to the couplings in
the fermionic and bosonic description by~\cite{ZZS}
\begin{equation}
\frac{8\pi}{\gamma} = 1 - \frac{\lambda}{\pi} = \frac{8\pi}{g^2} -
1\ .
\end{equation}
In particular, $\gamma = 8\pi$ and $g^2 = 4\pi$ is the free fermion
theory.

The Fourier transform of the kernel is fairly simple,
\begin{equation}
\tilde K(\omega) = \frac{\sinh\Bigl( \frac{\gamma\omega}{16} -
\frac{\pi\omega}{2} \Bigr)}
{2 \sinh \frac{\gamma\omega}{16} \cosh \frac{\pi\omega}{2} }\ .
\end{equation}
Thus
\begin{equation}
1 -\tilde K(0) = \frac{1}{2}\, \biggl( 1 + \frac{8\pi}{\gamma} \biggr) =
\frac{4\pi}{g^2}\ ,
\end{equation}
and so the energy and momentum densities are
\begin{equation}
{\mathcal E} + {\mathcal P}  = \frac{8\pi^2}{g^2}\, {\mathcal J}_R^2 
\ ,
\qquad
{\mathcal E} - {\mathcal P}  = \frac{8\pi^2}{g^2}\, {\mathcal J}_L^2 
\ . \label{eq:thirsol}
\end{equation}
On the other hand, canonical quantization gives
\begin{equation}
{\mathcal E} \pm {\mathcal P}  = \frac{1}{2g^2}\, (\phi' \pm g^2 
\Pi)^2 \ ,\qquad
\Pi = \dot \phi/g^2
  \label{eq:thircan}
\end{equation}
in the bosonic description of the massless theory.

There is an obvious correspondence between the integrable and canonical 
results~(\ref{eq:thirsol}) and (\ref{eq:thircan}).  However, it is 
slightly subtle to understand directly the relation between the quantum 
numbers ${\mathcal J}_R, {\mathcal J}_L$ of the integrable description 
and
those of the canonical description.  In the latter, there are two 
conserved charge densities.  The topological charge 
density $\frac{1}{2\pi}\phi'$ is the total fermion number density; this 
follows from our normalization of the field $\phi$  to periodicity 
$2\pi$.  Thus we identify ${\mathcal J}_R + {\mathcal J}_L = 
\frac{1}{2\pi} \phi'$.  The Noether density $2 \Pi$ is the chiral 
fermion number density; the normalization follows from the fact that $e^{\pm 
i\phi}$  are fermion bilinears.
We will see that this quantum number is more subtle to identify in the 
integrable case.  The conserved fermionic charges are
\begin{equation}
{\mathcal N}_{R,L} = \frac{1}{2}\,\biggl( \frac{1}{2\pi}\,\phi' \pm 2 \Pi 
\biggr)\ . \label{eq:dencan}
\end{equation}
We can now make two quick checks.  For the parity-symmetric state 
${\mathcal J}_R = {\mathcal J}_L = \frac{1}{4\pi} \phi' $, the chiral 
density $\Pi$ vanishes and in this case the energies~(\ref{eq:thirsol}) 
and (\ref{eq:thircan}) agree for all $g$.  For the free-fermion 
case $g^2 = 4\pi$, the fermions in the integrable and canonical 
descriptions are the same, 
\begin{equation}
{\mathcal J}_{R,L} = {\mathcal N}_{R,L}\qquad (g^2 = 4\pi)\ ,
\end{equation}
and with this the energies~(\ref{eq:thirsol}) and (\ref{eq:thircan}) 
match at the free fermion point.

To match the quantum numbers in general, let us start at the free 
fermion point and consider an adiabatic variation of $g$.  The 
densities~(\ref{eq:dencan}) are constructed from the topological and 
Noether densities without $g$-dependence, and so are invariant.  In 
terms of these, the canonical energy/momentum 
density~(\ref{eq:thircan}) is
\begin{equation}
{\mathcal E} \pm {\mathcal P}  = \frac{2\pi^2}{g^2} \biggl[  
({\mathcal N}_{R}
+ {\mathcal N}_{L}) \pm \frac{g^2}{4\pi}  ({\mathcal N}_{R}
- {\mathcal N}_{L}) \biggr]^2 \ . \label{eq:econ}
\end{equation}
To follow the quantum numbers in the integrable case, let us back up 
one step to the `undifferentiated' Bethe ansatz
\begin{equation}
m \sinh \theta + \frac{1}{2\pi i} \int_{-B_L}^{B_R}\!\! \ln S(\theta - 
\theta') \rho(\theta')
d\theta'
= \frac{2\pi n}{L}
\ ,\quad -B_L < \theta < B_R\ . \label{eq:bethezero}
\end{equation}
We have introduced a finite volume $L$; $n$ is an integer labeling the 
particle states.  Eq.~(\ref{eq:bethe}) is obtained from this by 
taking the difference for consecutive values of $n$.
All states in the range $-n_L < n < n_R$ are filled; the rapidity 
endpoints~$B_{R,L}$ are implicitly determined in terms of $n_{R,L}$.  
At the free fermion point we can immediately identify the number 
densities
\begin{equation}
{\mathcal N}_{R,L} = \frac{n_{R,L}}{L}\ .
\end{equation}
Both ${\mathcal N}_{R,L}$ and $n_{R,L}$ are adiabatically invariant, so 
this holds for all $g$.  Consider Eq.~(\ref{eq:bethezero}) in the 
conformal limit, taking
$\tilde \theta_R \to -\infty$:
\begin{eqnarray}
\frac{1}{2\pi i} \int_{-\tilde B_L}^{\infty}\!\! \ln {S_{}}_{LR} 
{\rho_{}}_L(\tilde\theta')
d\tilde\theta' + \frac{1}{2\pi i} \int_{-\infty}^{\tilde B_R}\! \ln 
{S_{}}_{RR}(-\infty) {\rho_{}}_R(\tilde\theta') d\theta'
&=& \frac{2\pi}{L}\, n_{\tilde\theta_R \to-\infty}
\nonumber\\[1mm]
\frac{1}{2\pi i} \int_{-\tilde B_L}^{\infty}\!\! \ln {S_{}}_{LL} (\infty)
{\rho_{}}_L(\tilde\theta')
d\tilde\theta' + \frac{1}{2\pi i} \int_{-\infty}^{\tilde B_R}\! \ln 
{S_{}}_{RL} {\rho_{}}_R(\tilde\theta') d\theta'
&=& \frac{2\pi}{L}\, n_{\tilde\theta_L \to\infty}
\ .\nonumber\\ \label{eq:ninf}
\end{eqnarray}
This determines the density ${\mathcal J}_R$, because the total number 
of filled right-moving states is
\begin{equation}
{L}{} {\mathcal J}_R = n_R - n_{\tilde\theta_R = -\infty}\ .
\end{equation}
Noting that
\begin{equation}
\frac{1}{2\pi i} \ln S_{LR} = - \frac{1}{2\pi i} \ln S_{RR}(-\infty) = 
\frac{1}{2}\, \tilde K(0) = \frac{1}{2}\,\biggl[1 - \frac{4\pi}{g^2} 
\biggr] \ ,
\end{equation}
the integrals~(\ref{eq:ninf}) just involve the total densities, giving
\begin{equation}
\frac{1}{2}\,\biggl[1 - \frac{4\pi}{g^2} \biggr] ({\mathcal J}_L - 
{\mathcal J}_R)
= {\mathcal N}_R - {\mathcal J}_R =  {\mathcal J}_L - {\mathcal N}_L\ ,
\end{equation}
where the last equality follows from a similar calculation for the 
left-movers.
Then
\begin{equation}
{\mathcal J}_{R,L}
= \frac{1}{2} \biggl[ ({\mathcal N}_R + {\mathcal N}_L) \pm
\frac{g^2}{4\pi}\, ({\mathcal N}_R - {\mathcal N}_L) \biggr]\ .
\label{eq:qnfin}
\end{equation}
With this identification of quantum numbers the energy/momentum 
densities~(\ref{eq:thirsol}) calculated using the integrable 
description do indeed reduce to those~(\ref{eq:econ}) obtained in the 
canonical description.

Again, the canonical currents~(\ref{eq:dencan}) are conserved under 
adiabatic variation of $g$, but the currents ${\mathcal J}_{R,L}$ 
are anomalous.  The Bethe equation~(\ref{eq:bethezero}) determines 
$\theta$ as a function of $n$ and $g$.
As we vary $g$ at fixed $n$, states move to the lower end of the 
right-moving spectrum and reappear on the upper end of the left-moving 
spectrum --- there is a spectral flow.

So far we have focused on the case that only a band of particle states is
filled, so that 
${\mathcal J}_{R}$ and ${\mathcal J}_{L}$ are positive.  However, it is clear
from Eq.~(\ref{eq:qnfin}) that as we vary $g$ one of these, say ${\mathcal
J}_{L}$, may go to zero.  What happens next is different in the nonconformal
theory, for arbitrarily small $m$, than in the conformal theory of interest:
the massless limit does not commute with adiabatic evolution.  In the massive
theory, when a left-moving particle state passes through zero rapidity it
becomes a right-moving particle state, so that after the last left-moving
particle has passed through zero we end up with a bounded interval of filled
positive rapidity particle states.
In the massless case, when a left-moving particle state passes through zero it
become an empty left-moving antiparticle state.  After the last filled
left-moving state passes through zero, empty left-moving particle states pass
through to become filled left-moving antiparticle states.  Thus, negative
values of ${\mathcal J}_{R,L}$ and ${\rho}_{R,L}$ signify filled antiparticle
states.

The derivation of the conformal Bethe ansatz equations~(\ref{eq:betheconf})
assumed particle states, but in fact these equations continue to hold.  The
particle-antiparticle reflection amplitude $S_R$ goes to zero at large rapidity
and the particle-antiparticle transmission amplitude $S_T$ goes to a
constant~\cite{ZZS}, so the right- and left-moving equations continue to
decouple.  Thus these equations apply for any signs of ${\mathcal J}_{R}$ and
${\mathcal J}_{L}$, as long as all right-movers are of the same type, and
similarly all left-movers.

\section{The \boldmath{$N \to 2$} Limit of the  \boldmath{$O(N)$} Coset Model}

In the classic study of $O(N)$-invariant S-matrices~\cite{ZZS}, the 
case $N=2$ required a separate treatment from $N>2$.  For example, the 
minimal $O(2)$ S-matrix~(\ref{eq:thirs}) contains the free parameter 
$\gamma$, while there is no free parameter for $N > 2$.  Thus the $N=2$ 
S-matrix cannot be thought of as a limit
from $N>2$.  However, we will argue in the next section that in order 
to treat the supergroup coset we need the analog of the $N \to 2$ limit 
of the S-matrix of the $O(N)$ sigma model.  We can think of this as 
corresponding to a different massive perturbation of the conformally 
invariant $O(2)$ model, turning on a nonzero $\beta$-function at $N = 
2+\epsilon$ rather than a nonzero fermion mass as in the Thirring 
description.  Of course, it is not clear a priori that this procedure 
is physically sensible, but we will try it and see.  We find that the 
$N \to 2$ limit of the Bethe ansatz appears to exist, but that it is 
more complicated than the conformal limits encountered thus far.

The sigma model action is
\begin{equation}
S = -\frac{1}{2g^2} \int d^2 x\, \partial_\mu \varphi^i \partial^\mu
\varphi^i \ ,\quad \varphi^i\varphi^i = 1\ ,\quad i = 1, \ldots, N\ .
\end{equation}
For $N=2$, $\varphi^1 + i \varphi^2 = e^{i\phi}$ gives the bosonic
action~(\ref{bosact}) at $m=0$.  The coupling $g$ runs for $N >2$ but this running turns
off in the limit, so by appropriately scaling the energy as we take $N \to 2$
we can obtain different fixed values of $g$.  The $\beta$-function is
\begin{equation}
\mu \,\frac{\partial g}{\partial\mu} = (N-2) F(g) \ ,\quad F(g) =
-\frac{g^3}{4\pi} - \frac{g^5}{8\pi^2} + \ldots\ . \label{eq:betaf}
\end{equation}
The coupling thus runs at a rate proportional to $N-2$,
\begin{equation}
(N-2) \ln \frac{\mu}{m} = \frac{2\pi }{g^2} + \ln g^2 + {\rm
const.} + \ldots  \ \equiv \chi(g)\ , \label{eq:pert}
\end{equation}
where $m$ is the dynamically generated mass scale.
Identifying $\mu \sim E \sim me^{|\theta|}$, we see that when we hold $E$ and $g$ fixed as $N \to 2$, the dynamical mass $m$ goes to zero, and also we must hold fixed $|\theta| - \chi(g)/(N-2)$.  That is, we focus on a rapidity region where the  coupling takes a specified value $g$ in the limit.  

The S-matrix for the $O(N)$ sigma model decomposes into three terms
\begin{eqnarray}
| k\,\theta, l\, \theta'; {\rm in} \rangle &=& S_{kl,ij}(\theta - \theta') 
| i\,\theta, j\, \theta'; {\rm out} \rangle\ ,
\nonumber\\[2mm]
S_{kl,ij}(\theta ) &=& \delta_{ij} \delta_{kl} \sigma^+_1(\theta) 
+ \delta_{ik} \delta_{jl} \sigma^+_2(\theta) 
+ \delta_{il} \delta_{jk} \sigma^+_3(\theta) \ ,\ \ \label{eq:onsmat}
\end{eqnarray}
where $\sigma^+_{1,2,3}(\theta-\theta')$ are given in Ref.~\cite{ZZS}.
As in the limiting process of Sec.~1, RR and LL scattering involve finite
differences in rapidity while RL scattering involves rapidity differences that
diverge in the limit, as $1/(N-2)$.  The $O(N)$ sigma model S-matrix for
same-charge scattering is 
$S = \sigma_2^+ + \sigma_3^+$, which is
\begin{equation}
S(\theta)= \frac{\Gamma\Bigl( \frac{1}{N-2} - \frac{i\theta}{2\pi} \Bigr)
\Gamma\Bigl( \frac{1}{N-2} + \frac{1}{2} + \frac{i\theta}{2\pi} \Bigr)
\Gamma\Bigl( \frac{1}{2} - \frac{i\theta}{2\pi} \Bigr)
\Gamma\Bigl(  \frac{i\theta}{2\pi} \Bigr)}{\Gamma\Bigl( \frac{1}{N-2} +
\frac{i\theta}{2\pi} \Bigr)
\Gamma\Bigl( \frac{1}{N-2} + \frac{1}{2} - \frac{i\theta}{2\pi} \Bigr)
\Gamma\Bigl( \frac{1}{2} + \frac{i\theta}{2\pi} \Bigr)
\Gamma\Bigl(  \frac{-i\theta}{2\pi} \Bigr)}\ .
\end{equation} 
The limit relevant to the LL and RR S-matrices is taken with fixed rapidity,
\begin{equation}
\lim_{N\to 2} S(\theta) = \frac{\Gamma\Bigl( \frac{1}{2} - \frac{i\theta}{2\pi}
\Bigr)
\Gamma\Bigl(  \frac{i\theta}{2\pi} \Bigr)}{
\Gamma\Bigl( \frac{1}{2} + \frac{i\theta}{2\pi} \Bigr)
\Gamma\Bigl(  \frac{-i\theta}{2\pi} \Bigr)} \equiv S_{\rm I}(\theta)\ .
\end{equation}
The limit relevant to the LR S-matrix is taken with rapidity proportional to $1/(N-2)$, because the right- and left-movers are localized near $\theta = \pm\chi(g)/(N-2)$.  Thus we define
\begin{equation}
\lim_{N\to 2} S\bigl( \zeta/[N-2]\bigr) = \biggl( \frac{2\pi + i \zeta}{2\pi -
i\zeta} \biggr)^{\! 1/2}
e^{i \pi\, {\rm sign}(\zeta)/2} \equiv S_{\rm II}(\zeta)\ ,
\end{equation}
and $S_{LR} = S_{\rm II}(2\chi(g))$.  The coupling does not appear in $S_{LL}$ and $S_{RR}$, but does appear in $S_{LR}$.

Now, however, we encounter an interesting complication.  In the Thirring model
we had
\begin{equation}
S_{LR} = S_{LL}(\tilde\theta \to \infty)\ ,\quad S_{RL} = S_{RR}(\tilde\theta
\to -\infty)\ ,
\label{eq:rlmatch}
\end{equation}
so the particle numbers in Eq.~(\ref{eq:ninf}) satisfy $n_{\tilde\theta_R
\to-\infty} = n_{\tilde\theta_L \to\infty}$.  That is, there are no missing
$n$'s between the right- and left-movers.  In the present case, this cannot
hold in general, because $S_{RR}$ and $S_{LL}$ do not depend
on the coupling while $S_{RL}$ and $S_{LR}$ do.  Thus, there is a range of $n$
that correspond to what we will call `zero mode' states, in the large rapidity
regime between the right- and left-movers.  If we solve the Bethe ansatz for
$N>2$ and then take the limit, the rapidity distribution must approach such a form.
 Thus we generalize the earlier conformal limit~(\ref{eq:conlim})
to\,\footnote{\,We can divide the rapidity range so that right-movers have $\theta
> \chi - \epsilon^{-1}$, left-movers have $\theta
< -\chi + \epsilon^{-1}$, and zero modes are in between.  As long as $\epsilon$
goes to zero as $N\to 2$ but does so more slowly than $N-2$ itself (e.g.
$\epsilon = \sqrt{N-2}$), one gets the indicated ranges for $\tilde\theta_R$,
$\tilde\theta_L$, and $\zeta$.}
\begin{eqnarray}
\rho_R(\tilde\theta_R) &=& \lim_{N\to 2} \rho\bigl(\tilde\theta_R + \chi(g)/[N-2]
\bigr)\ ,
\quad -\infty < \tilde\theta_R < \tilde B_R\ ,
\nonumber\\[1mm]
\rho_L(\tilde\theta_L) &=& \lim_{N\to 2} \rho\bigl(\tilde\theta_L  - \chi(g)/[N-2]
\bigr)\ ,
\quad -\tilde B_L < \tilde\theta_R < \infty\ ,
\nonumber\\[1mm]
\rho_0(\zeta) &=& \lim_{N\to 2} \frac{1}{N-2}\, \rho\bigl( \zeta/[N-2]\bigr)\
,\quad
-\chi(g) < \zeta < \chi(g)\ .
\end{eqnarray}
Because the zero modes occupy a range of $\theta$ of order $(N-2)^{-1}$, their
density must be of order $N-2$, and so we have included a compensating factor
in the definition of $\rho_0$.

The right- and left-moving Bethe ansatz equations are exactly as in
Eq.~(\ref{eq:betheconf}),
using $K_{\rm I}$ constructed from $S_{\rm I}$.  In particular, the zero-modes
do not enter into these equations because the rapidity difference is large and
$\partial_\theta S_{\rm II}$ is of order $N-2$.  To write the zero mode
equations we define
\begin{eqnarray}
\frac{1}{2\pi i}\, \partial_\zeta \ln S_{\rm II}(\zeta) &=&  K_{\rm II}(\zeta)
= \frac{1}{2}\, \delta(\zeta) + k(\zeta)\ ,
\nonumber\\[1mm]
k(\zeta) &=& \frac{1}{ 4\pi^2+\zeta^2}\ .
\end{eqnarray}
Then
\begin{eqnarray}
\int_{-\chi}^{\chi} k(\zeta - \zeta') \rho_0(\zeta')
d\zeta' + 2\pi  \bigl[ k(\zeta -\chi) {\mathcal Q}_R + k(\zeta +\chi) {\mathcal
Q}_L \bigr] 
&=& \frac{1}{2}\, \rho_0(\zeta)\ ,
\nonumber\\
-\chi < \zeta &<& \chi \ .
\label{eq:zero}
\end{eqnarray}
We use ${\mathcal Q}_{R,L}$ here to distinguish these from the densities
${\mathcal J}_{R,L}$ of the fermionic description.  The coupling $g$ now enters into the Bethe ansatz equations only through the implicit $g$-dependence of the rapidity range $\chi$.

The Bethe ansatz equations for $\rho_R$ and for $\rho_L$ separate from the
other components of $\rho$, while the total ${\mathcal Q}_R$ and ${\mathcal
Q}_L$ give rise to inhomogeneous terms in the $\rho_0$ equation.  The zero
modes do feed back into the undifferentiated Bethe
equation~(\ref{eq:bethezero}) which determines the total ${\mathcal Q}_R$ and
${\mathcal Q}_L$.
The energy of the zero modes is exponentially small in the limit, so the energy
and momentum come only from the right- and left-movers as in
Eqs.~(\ref{eq:ep1},\,\ref{eq:ep2}):
\begin{equation}
{\mathcal E} + {\mathcal P}  = {\pi} {\mathcal Q}_R^2  \ ,
\qquad
{\mathcal E} - {\mathcal P}  = {\pi} {\mathcal Q}_L^2  \ . \label{eq:n2en}
\end{equation}
The zero modes affect the energy indirectly because they enter into the
determination of ${\mathcal Q}_R$ and ${\mathcal Q}_L$.

As in the previous section, the Bethe ansatz has been derived by taking the
limit of a state with particles only, but it can be extended to negative
rapidity densities.  Using the expressions in Ref.~\cite{ZZS}, the
particle-antiparticle
reflection amplitude $S_R = \sigma^+_1 + \sigma^+_3$ vanishes for rapidities of
order $1/(N-2)$,
while the particle-antiparticle transmission amplitude $S_T = \sigma^+_1 +
\sigma^+_2$ gives a kernel which is equal to $-k(\zeta)$.  Therefore we can use
the Bethe ansatz equations
(\ref{eq:betheconf},\,\ref{eq:zero}) freely for positive and negative densities
as long as particle and antiparticles are separated by rapidities of order
$1/(N-2)$.  That is, all the right-movers must be of one type, and all the
left-movers similarly, but the zero mode density may have a sign that changes
as a function of rapidity, since the typical rapidity difference for the zero
modes is of order $1/(N-2)$.

The zero modes represent a substantial complication, because their rapidity
support is bounded in both directions.  In fact, the kernel $k$ is essentially
the same as that for the nonconformal $O(3)$ model, and so the zero mode
equation can only be solved as a series in $\chi$ or in $1/\chi$~\cite{has1}.
We see from the perturbative calculation~(\ref{eq:pert}) that the expansion in
$1/\chi$ corresponds to small $g$ in the nonlinear sigma model.  In the other
limit, $\chi \to 0$, the zero modes disappear.  Taking the case that ${\mathcal
Q}_R ={\mathcal Q}_L \equiv {\mathcal Q}/2 $, we have ${\mathcal E} = \pi
{\mathcal Q}^2/4$.  On the other hand, since the zero modes carry no charge in
this limit we can identify ${\mathcal Q}$ with the Noether charge $\Pi$, in
terms of which ${\mathcal E} = g^2 \Pi^2 / 2$.  Thus we identify $g^2 = \pi/2$
with $\chi \to 0$.  The $N \to 2$ limit therefore can reach the range of
couplings
\begin{equation}
0 < g^2 \leq \frac{\pi}{2}\ ,\qquad \infty > \chi \geq 0\ .
\end{equation}
It is curious that we cannot reach all values of the coupling $g$; perhaps
there is some extension or continuation of our construction that makes it
possible to do so.

The expansion for small $\chi$ is straightforward because the integral term in
the zero mode equation~(\ref{eq:zero}) is small in this limit and the equation
can be solved iteratively; also the kernel $k$ can be expanded in $\zeta$,
which is of order $\chi$.  One readily obtains
\begin{equation}
{\mathcal Q}_0 = ({\mathcal Q}_R + {\mathcal Q}_L)\Biggl\{
\frac{\chi}{\pi^2} + \frac{\chi^2}{\pi^4}
+ \biggl[1 - \frac{\pi^2}{3}  \biggr] \frac{\chi^3}{\pi^6} 
+ \biggl[1 - \frac{\pi^2}{2} \biggr] \frac{\chi^4}{\pi^8}+
\ldots \Biggr\}\ .
\end{equation}
Identifying the total Noether charge ${\mathcal Q}_0  + {\mathcal Q}_R +
{\mathcal Q}_L = \Pi$, we can also write
\begin{equation}
{\mathcal Q}_R + {\mathcal Q}_L = 
\Pi \biggl\{1 
- \frac{\chi}{\pi^2} 
+ \frac{\chi^3}{3 \pi^4} 
- \frac{\chi^4}{6 \pi^6} +
\ldots \biggr\}\ .
\end{equation}
If we take again the state with ${\mathcal Q}_R ={\mathcal Q}_L = 
{\mathcal Q}/2 $, matching the energies
$\pi {\mathcal Q}^2/4 = g^2 \Pi^2 / 2$
 as in the previous paragraph gives
\begin{equation}
g^2 = \frac{\pi}{2}  \biggl\{1
- 2 \,\frac{\chi}{\pi^2} 
+ \frac{\chi^2}{\pi^4} 
+ \frac{2\chi^3}{3\pi^4} 
- \frac{\chi^4}{ \pi^6} +
\ldots \biggr\}\ . \label{smallc}
\end{equation}
Thus we obtain the functional relationship between the coupling (radius) in the
$O(2)$ theory and the parameter $\chi$ which governs the $N\to 2$ limit of the
rapidity difference between the right- and left-movers, in the neighborhood of
$g^2 \sim \pi/2$.

For states with 
${\mathcal Q}_R \neq {\mathcal Q}_L$ we could again use the undifferentiated
equation~(\ref{eq:bethezero}) in order to identify the quantum numbers, but a
simple shortcut is to notice that the momentum density ${\mathcal P} = \pi
({\mathcal Q}_R^2 - {\mathcal Q}_L^2)/2 = \Pi\phi'$ is adiabatically invariant.
 We then have
\begin{eqnarray}
\hspace{-5mm}{\mathcal Q}_R - {\mathcal Q}_L &=& \frac{2\phi'}{\pi} \,\frac{\Pi}{{\mathcal
Q}_R + {\mathcal Q}_L} = \frac{2\phi'}{\pi}\, \biggl[ 1 + \frac{{\mathcal
Q}_0}{{\mathcal Q}_R + {\mathcal Q}_L} \biggr] \nonumber\\[1mm]
\hspace{-5mm}&=&
\frac{2\phi'}{\pi} 
\Biggl\{1+
\frac{\chi}{\pi^2} + \frac{\chi^2}{\pi^4}
+ \biggl[1\! - \!\frac{\pi^2}{3}  \biggr] \frac{\chi^3}{\pi^6} 
+ \biggl[1\! -\! \frac{\pi^2}{2} \biggr] \frac{\chi^4}{\pi^8}+
\ldots \Biggr\}\ .
\end{eqnarray}

The $1/\chi$ expansion is more involved, and the conformal limit appears no
simpler than the general case.  We therefore simply take the $N \to 2$ limit of
the known $O(N)$ result.  Using Eqs.\,(13,\,22,\,23) of Ref.~\cite{has2}, with $\Pi
= \partial f/\partial h$ and $\chi = (N-2)B$, gives
\begin{equation}
\frac{\mathcal E}{\Pi^2} = \frac{\pi}{\chi} - \frac{\pi}{\chi^2} \ln 8\chi +
\ldots\ .
\end{equation}
Equating this to the canonical result~$g^2/2$ and solving for $\chi$ leads to
\begin{equation}
\chi = \frac{2\pi}{g^2} + \ln g^2 + \ln\frac{4}{\pi} + \ldots\ . \label{largec}
\end{equation}
In particular, this reproduces the two-loop result~(\ref{eq:pert}).

The results~(\ref{smallc},\,\ref{largec}) just give the relation between the
parameter $g$ of the canonical description and the parameter $\chi$ of the
integrable description.  Once this relation is known, one can use the
integrable description to calculate physical quantities, such as the spectrum
of excitations.  Of course, in this case the canonical description is vastly
simpler.

\section{The \boldmath{$OSp(2\!+\!2M|2M)$} Coset Model}

In Sec.~3 we solved free field theory in a difficult way, and then in Sec.~4 we
solved it in an even more difficult way.  We can now take these efforts and
apply them rather directly to a conformal theory which is not free, and not
solvable by the usual methods of chiral algebra.
The $OSp(N+2M|2M)$ coset model has the action
\begin{equation}
S = -\frac{1}{2g^2} \int d^2 x\, J_{ij} \partial_\mu \varphi^i \partial^\mu
\varphi^j \ ,\quad  J_{ij} \varphi^i\varphi^j = 1\ .
\end{equation}
The components $\varphi^i$ have statistics
\begin{eqnarray}
{\rm commuting:}&&\ 1 \leq i \leq N+2M\ ,\nonumber\\[1mm]
 {\rm anticommuting:}&&\ N + 2M + 1 \leq i \leq N + 4M\ ,
\end{eqnarray}
and $J_{ij}$ is
\begin{equation}
J = \left[
\begin{array}{ccc}
I_{N+2M}  &  0 & 0  \\
 0 & 0  & -I_M  \\
0  & ~I_M  & 0  
\end{array}
\right]\ .
\end{equation}

Consider an amplitude in which only the first $N$ bosonic fields are present in
the external states and operators.  The remaining $2M$ bosonic fields and the
$2M$ fermionic fields appear only in loops, and by drawing the graphs in
single-line notation (or introducing an auxiliary field to make the integral
over $\varphi$ gaussian) it becomes evident that these $M$-dependent
contributions cancel.  Thus these amplitudes are independent of $M$, and are
the same in the supergroup model as in the bosonic $O(N)$ sigma
model~\cite{ParSour,Weg}.  In particular, the S-matrix for states involving
only the first $N$ bosonic components is identical to the $O(N)$
result~(\ref{eq:onsmat}).  But then there is a unique $OSp(N+2M|2M)$-invariant
extension~\cite{SWK},
\begin{eqnarray}
| k\,\theta, l\, \theta'; {\rm in} \rangle &=& S_{kl,ij}(\theta - \theta') 
| i\,\theta, j\, \theta'; {\rm out} \rangle
\nonumber\\[1mm]
S_{kl,ij}(\theta ) &=& J^{\vphantom 1}_{ij} J^{-1}_{kl} \sigma^+_1(\theta) 
+ \delta_{ik} \delta_{jl} \sigma^+_2(\theta) 
+ (-1)^{p_i + p_j} \delta_{il} \delta_{jk} \sigma^+_3(\theta) \ ,\ \
\label{eq:ospS}
\end{eqnarray}
where $p_i$ is 0 when $\varphi^i$ is bosonic and 1 when it is fermionic.
This S-matrix is not unitary, but preserves an indefinite inner product built
from $J$.

We can now take the $N \to 2$ limit as before, and in this way obtain the Bethe
ansatz for the conformally invariant $OSp(2\!+\!2M|2M)$ coset.  The finite field
calculations in Ref.~\cite{has1,has2} and in Sec.~4 involve only one $O(2)$
charge and so lift directly to the $OSp(2\!+\!2M|2M)$ coset~\cite{SWK} --- for
states with only a single $O(2) \subset O(2\!+\!2M)$ charge the energy reduces to
that of the $O(2)$ theory and so can be calculated in free field theory.  Now,
however, we can go on to consider more general states, having charges in more
than one $O(2)$ subgroup of $OSp(2\!+\!2M|2M)$.  We will leave the detailed study
of these states for future work.

Why not can we not lift the simpler massless Thirring S-matrix of Sec.~3 to
$OSp(2\!+\!2M|2M)$?
The difficulty is that the Thirring fermions have no simple transformation
property under $OSp(2\!+\!2M|2M)$.  They are spinors under the first $O(2) \subset
O(2\!+\!2M)$ but are neutral under the commuting $O(2)$'s, so they do not even lift
to a spinor representation of $OSp(2\!+\!2M|2M)$.
Thus the Thirring description does not seem useful for the
supercoset.\footnote{\,R. Roiban has independently pointed out another
difficulty.  If we simply lift the Thirring S-matrix as in Eq.~(\ref{eq:ospS}),
treating the Thirring fermions as vectors of $OSp(2\!+\!2M|2M)$, it does not
satisfy the Yang-Baxter equation.  The existence of an
S-matrix~(\ref{eq:thirs}) containing a free parameter $\gamma$ depends on
identities that are special to $O(2)$ and do not lift to $OSp(2\!+\!2M|2M)$.}

\section{Discussion}

We have found (Sec.~2,\,3) that for some integrable theories, the conformal
limit leads to simplifications in the Bethe ansatz.  For others (Sec.~4) it
does not.  Unfortunately, the supergroup coset $OSp(2\!+\!2M|2M)$ appears to be of
the latter type.  It is conceivable that there is a simpler description of this
model; different massive perturbations of a given conformal theory define
different bases of states and so different S-matrices.  However, we suspect
that in the present case there may simply be a certain irreducible complexity
to the integral equations that must be solved.

Since this work is ultimately directed at a better understanding of the AdS/CFT
duality, let us list the steps that would be needed to reach this goal.  First,
one must find the S-matrix having the appropriate symmetry, for example
$PSU(2,2|4)$, and the appropriate degrees of freedom.  This S-matrix is
likely to be similar in complexity to the $OSp(2\!+\!2M|2M)$ S-matrix.  In our case
we were aided by having a family of nonconformal theories whose limit we could
take; perhaps given this example one can determine the S-matrix for other
conformal supergroup theories directly.  Second, one must understand how the
BRST ghost degrees of freedom of the world-sheet theory enter into the
integrable description, and how the BRST charge acts on the states in the
integrable description.

Third, to obtain a complete integrable description of the spectrum one must go
beyond the large-field case and understand finite volume effects.  The problem
is that continuum S-matrices such as those that we have used are defined with
reference to the infinite-volume vacuum.  Putting the system in a finite volume
changes the vacuum (for example there is a Casimir energy) and so changes the
excitation energies and the S-matrix.\footnote{
One gets wrong answers if one simply ignores this and forges ahead.  The simplest example is the state of a single fermion with $n$ units of right-moving momentum.
The naive Bethe ansatz would give energy $2\pi n/L$ independent of $g$ (this is trivial, since the S-matrix does not enter).  On the other hand, the CFT calculation using has explicit $g$-dependent terms involving the fermion charges (or the momentum and winding, in bosonized form).}
This poses a great difficulty, and so
for the most part finite volume energies are understood only for the ground
state~\cite{Ztba} and for twisted sector ground states~\cite{Mart,KM,Fen},
via a world-sheet space-time duality and the Thermodynamic Bethe Ansatz (Refs.~\cite{BLZ2,DT,FMQR} are able to go to a larger set of states).  The other approach to finite volume energies is to
find a discretized version of the integrable theory, for which there is a
trivial ferromagnetic vacuum, and build the theory around this vacuum (for work
on discrete supercoset models see ref.~\cite{RS}).  We had hoped that the
large-charge states such as those that we are considering could play the role
of such a ferromagnetic vacuum but directly in the continuum theory.  However,
these still have low-lying excitations and long-range correlations, and so they
are not as simple as would be needed.

For the $OSp(2+2M|2M)$, or its continuation to $OSp(2,2M|2M)$ there is the interesting question as to whether the large-curvature limit $g^2 \to \infty$ has any simple dual description; in the AdS/CFT case this is the limit where the dual field theory becomes weakly coupled.
Finally, we believe that our most interesting result is the nature of the
$N\to 2$ limit of the $O(N)$ and $OSp(N+2M|2M)$ theories: this limit seems to
be sensible, but has a nontrivial zero mode sector in addition to the right-
and left-movers.  Some of the features that we have found may arise in other
approaches to the integrability of supergroup models.

\section*{Acknowledgments}

We would like to thank O. DeWolfe, T. Erler, R. Roiban, and A. Volovich for many
discussions, and also O. DeWolfe for detailed comments on the manuscript.  We would also like to thank P. Dorey, D. Fioravanti, P. Hasenfratz, A. Ludwig, F.
Niedermayer and M. Staudacher for communications.
This work was supported by National Science Foundation
grants PHY99-07949 and PHY00-98395.  The work of N.M. was also supported by a
National Defense Science and Engineering Graduate Fellowship.

\end{document}